\documentclass{iopart}              

\usepackage{iopams}
\usepackage{graphicx}

\begin{document}

\title[]{Magnetoresistance, specific heat and magnetocaloric effect of
equiatomic rare-earth transition-metal magnesium compounds}

\author{H~Hartmann$^1$, K~Berggold$^1$, S~Jodlauk$^1$, I~Klassen$^1$,
K~Kordonis$^1$, T~Fickenscher$^2$, R~P\"{o}ttgen$^2$, A~Freimuth$^1$
and T~Lorenz$^{1,3}$\footnote[0]{$^3$ Author to whom
correspondence should be addressed (lorenz@ph2.uni-koeln.de).} }


\address{$^1$ II.\,Physikalisches Institut, Universit\"{a}t zu K\"{o}ln, Z\"{u}lpicher Str.
77, 50937 K\"{o}ln, Germany}

\address{$^2$ Institut f\"{u}r Anorganische und Analytische Chemie,
Westf\"{a}lische Wilhelms-Universit\"{a}t M\"{u}nster, Correnstrasse 30, 48149
M\"{u}nster, Germany}

\begin{abstract}
We present a study of the magnetoresistance, the specific heat
and the magnetocaloric effect of equiatomic $RET$Mg intermetallics
with $RE = {\rm La}$, Eu, Gd, Yb and $T = {\rm Ag}$, Au and of
GdAuIn. Depending on the composition these compounds are
paramagnetic ($RE = {\rm La}$, Yb) or they order either ferro- or
antiferromagnetically with transition temperatures ranging from
about 13 to 81\,K. All of them are metallic, but the resistivity
varies over 3 orders of magnitude. The magnetic order causes a
strong decrease of the resistivity and around the ordering
temperature we find pronounced magnetoresistance effects. The
magnetic ordering also leads to well-defined anomalies in the
specific heat. An analysis of the entropy change leads to the
conclusions that generally the magnetic transition can be
described by an ordering of localized $S=7/2$ moments arising
from the half-filled $4f^7$ shells of Eu$^{2+}$ or Gd$^{3+}$.
However, for GdAgMg we find clear evidence for two phase
transitions indicating that the magnetic ordering sets in
partially below about 125\,K and is completed via an almost
first-order transition at 39\,K. The magnetocaloric effect is
weak for the antiferromagnets and rather pronounced for the
ferromagnets for low magnetic fields around the zero-field Curie
temperature.
\end{abstract}




\section{Introduction}

The equiatomic rare-earth($RE$) transition-metal($T$) magnesium
compounds $RET$Mg have intensively been investigated in recent
years with respect to their crystal chemistry and to a basic
characterization of their magnetic
properties.\cite{P1,P2,P3,P4,P5,P6,P7,P8,P9,P10,P11,P12} With the
trivalent rare earth elements, the RE$T$Mg intermetallics adopt
the hexagonal ZrNiAl structure,\cite{P13,P14,P15} while those
with a divalent rare earth metal, i.\,e.\ europium and ytterbium,
crystallize in the orthorhombic TiNiSi type structure.\cite{P16}
Interesting magnetic properties have been observed for several of
the $RET$Mg compounds. To give some examples, long-range magnetic
ordering has been observed for Ce$T$Mg ($T = {\rm Pd}$, Pt, Au)
at 2.1, 3.6, and 2.0\,K, respectively.\cite{P9} EuAgMg and EuAuMg
order ferromagnetically at relatively high Curie temperatures of
22 and 37\,K.\cite{P6} Similar behaviour was observed for the
gadolinium-based compounds Gd$T$Mg ($T = {\rm Pd}$, Ag, Pt),
which show ferromagnetic order at 96, 39, and 98\,K for $T = {\rm
Pd}$, Ag, Pt, respectively.\cite{P10} In contrast, GdAuIn and
GdAuMg order antiferromagnetically with N\'{e}el temperatures of 13
and 81\,K, respectively.\cite{P12} Not much is known, however,
about the transport and thermodynamic properties of these
compounds, in particular about their magnetic field dependencies.
In this report we present resistivity measurements and specific
heat data  for some $RE$AgMg and $RE$AuMg compounds and for
GdAuIn in the temperature range from about 2 to 300\,K and in
external magnetic fields up to 14\,T. We have also extended the
magnetization measurements up to 14\,T, which we use to calculate
the magnetocaloric effect.

\section{Experimental}


Starting materials for the preparation of the intermetallic
compounds $RET$Mg and GdAuIn were sublimed ingots of the rare
earth elements (Johnson-Matthey, Chempur or Kelpin, $> 99.9$\,\%),
silver and gold wire (diameter 1 mm, Degussa- H\"{u}ls, $>
99.9$\,\%), a magnesium rod (Johnson Matthey, diameter 16 mm,
$>99.5$\,\%), and indium tear drops (Johnson Matthey, 99.9\%).
GdAuIn was prepared by arc-melting\,\cite{P18} of the elements
under an atmosphere of $\simeq 600$\,mbar argon. The argon was
purified over titanium sponge (970\,K), silica gel, and molecular
sieves prior to use. The arc-melted button was remelted three
times to ensure homogeneity. Due to the low boiling point and the
high vapor pressure, the magnesium containing samples cannot be
obtained via a simple arc-melting procedure, where a significant
weight loss would occur through evaporation of magnesium. All
these compounds were prepared in sealed high-melting metal tubes.
The elements were mixed in the ideal 1:1:1 atomic ratios and
sealed in small (tube size $\simeq 1$\,cm$^3$) niobium or
tantalum tubes under an argon pressure of $\simeq 800$\,mbar.
These tubes were then put in a water-cooled sample
chamber\,\cite{P19} of a high-frequency furnace (H\"{u}ttinger
Elektronik, Freiburg, TIG 1.5/300). The samples were first
rapidly heated to $\simeq 1370$\,K and subsequently annealed at
$\simeq 870$\,K for another two hours. The temperature was
controlled through a Sensor Therm Metis MS09 pyrometer with an
accuracy of $\pm 30$\,K. The samples could be separated from the
tubes by mechanical fragmentation. No reaction with the metal
tubes was observed. The polycrystalline samples are light grey
with metallic lustre. For more details concerning the synthesis
conditions we refer to the original literature.


The purity of the samples was checked through Guinier powder
patterns using CuK$\alpha_1$ radiation and $\alpha$-quartz as an internal
standard. The Guinier camera was equipped with an image plate
system (Fujifilm, BAS-1800). The experimental patterns were
compared with calculated ones,\cite{P20} using the atomic
positions from the previous single crystal
studies.\cite{P4,P5,P6,P7,P8,P9,P10,P11,P12} All samples were
pure phases on the level of X-ray powder diffraction.

For the measurements of resistivity, magnetization and specific heat
we have polished larger polycrystalline pieces of irregular shape
to a rectangular shape of typical dimensions of
about $2 \times 1 \times 1$\,mm$^3$. The specific
heat and the magnetization have been studied in a Physical
Property Measurement System (PPMS, Quantum Design) using the
relaxation time method and a vibrating sample magnetometer,
respectively. The resistivity has been measured by a standard DC four-probe
technique. All these measurements have been performed in the
temperature range from about 2 to 300\,K and in magnetic fields
up to 14\,T.


\section{Results and discussion}

\subsection{Crystal chemistry}

The crystal structures of the $RET$Mg intermetallics depend on the
valence of the rare earth element. With the trivalent rare earth
metals, the $RET$Mg compounds adopt the hexagonal ZrNiAl,
type,\cite{P13,P14,P15} while the europium and ytterbium
compounds contain Eu$^{\rm II}$ and Yb$^{\rm II}$ and they adopt
the orthorhombic TiNiSi type.\cite{P16} In both structure types,
the transition metal and magnesium atoms build up a
three-dimensional [$T$Mg] network in which the rare earth atoms
fill distorted hexagonal channels. GdAuIn also crystallizes with
the ZrNiAl structure. For more details on the crystal chemistry
and chemical bonding of these $RET$Mg intermetallics we refer to
the original literature.\cite{P4,P5,P6,P7,P8,P9,P10,P11,P12,P17}

\subsection{Resistivity}

\begin{figure}[t]
\begin{center}
\includegraphics[width=0.7\textwidth]{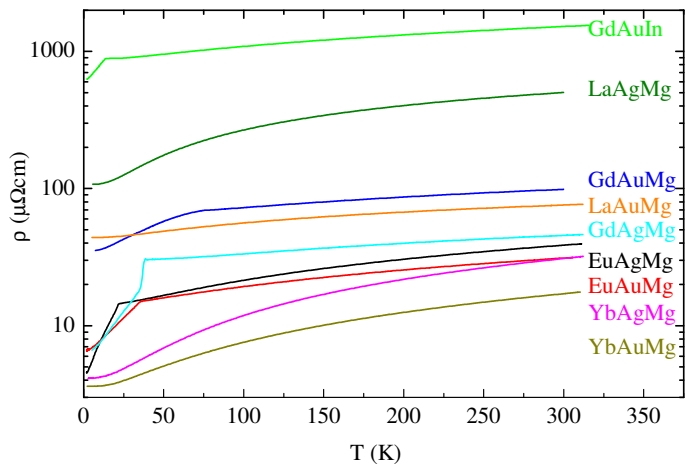}
\end{center}
\caption{\label{rho} Resistivity of $RET$Mg with $RE={\rm Yb}$,
Eu, Gd, or La and $T={\rm Au}$ or Ag and of GdAuIn.}
\end{figure}

Figure\,\ref{rho} gives an overview of the resistivity $\rho$ of
the different compounds. For all samples the temperature
dependence of $\rho$ is metallic. However, the absolute values are
strongly different, ranging from a few $\mu\Omega cm$ in YbAuMg up
to $m\Omega cm$ in GdAuIn. We find the following trends. The
$RET$Mg samples with divalent Yb and Eu have lower $\rho$ values
than the compounds with trivalent La and Gd. This indicates that
the different $\rho$ is related to the different structure of the
divalent (TiNiSi type) and of the trivalent (ZrNiAl type)
$RE$-based compounds. GdAuIn also adopts the ZrNiAl structure,
but it has a significantly larger $\rho$ than GdAuMg. Possibly
this is related to the different valences of In (trivalent) and Mg
(divalent). Combining this observation with the different
resistivities of the trivalent and the divalent $RE$-based
compounds, one observes that $\rho$ increases with increasing
number of valence electrons for the entire series $RETX$.
Moreover, almost all $RE$AuMg samples have lower $\rho$ values
than the corresponding $RE$AgMg samples (the only exception is
Gd$T$Mg). LDA+U band-structure calculations and photoemission
studies are currently underway in order to check whether these
trends correlate with systematic changes in the band structure for
the different compounds.

\begin{figure}[t]
\begin{center}
\includegraphics[width=\textwidth]{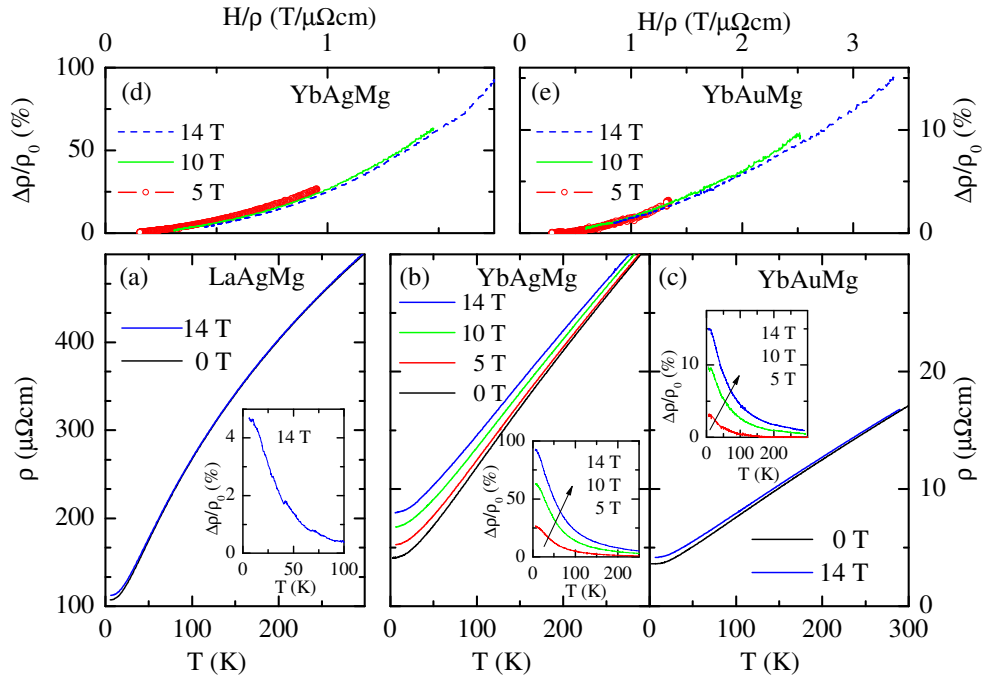}
\end{center}
\caption{\label{rhopm} Resistivity of paramagnetic $RET$Mg for
different magnetic fields [panels (a)--(c)]. The respective insets
show the magnetoresistance $\Delta \rho/\rho_0=
[\rho(H)-\rho(H=0)]/\rho(H=0)$ (increasing field strength is
indicated by the arrows). Panels (c) and (d) show that the
magnetoresistance curves for different field strengths follow a
single line if $\Delta \rho/\rho_0$ is plotted as a function of
$H/\rho_0$ as it is expected from Kohler's rule.}
\end{figure}

There are pronounced kinks in $\rho (T)$ of the Gd- and Eu-based
compounds. These kinks arise from the ferro- or antiferromagnetic
ordering of the $4f$ moments of Gd and Eu. In contrast, the La-
and Yb-based materials show continuous $\rho (T)$ curves without
any anomalies reflecting that the latter materials are
paramagnetic down to the lowest temperature. In
figure\,\ref{rhopm} we show the resistance of the paramagnets for
different magnetic fields. For all samples we find a systematic
increase of $\rho$ with increasing field. This increase becomes
the more pronounced the lower the temperature is. In the insets
of figure\,\ref{rhopm} we present the normalized
magnetoresistance $\Delta \rho/\rho_0=
[\rho(H)-\rho(H=0)]/\rho(H=0)$. For YbAgMg $\Delta \rho/\rho_0$
reaches more than 90\,\% at about 5\,K and a field of 14\,T,
whereas only about 15\,\% is obtained for YbAuMg. In LaAgMg the
absolute change $\Delta \rho$ is comparable to that of YbAgMg,
but the relative change $\Delta \rho/\rho_0$ is reduced by about a
factor of 20 due to the larger $\rho_0$ of LaAgMg. The
field-induced increase of $\rho$ is a consequence of the
additional scattering of the charge carriers due to the Lorentz
force as it is observed in many metals.\cite{ziman} Within
semiclassical transport theory one expects the so-called Kohler's
rule to hold, which predicts that $\Delta \rho/\rho_0$ follows a
universal function $ f\left(H/\rho_0\right)$ over the entire
field and temperature range. As shown in the upper panels of
figure\,\ref{rhopm}, this is rather well fulfilled. For both
Yb-based samples $f(x)$ is roughly proportional to
$x^2$.\cite{remarkLa}

\begin{figure}[t]
\begin{center}
\includegraphics[width=\textwidth]{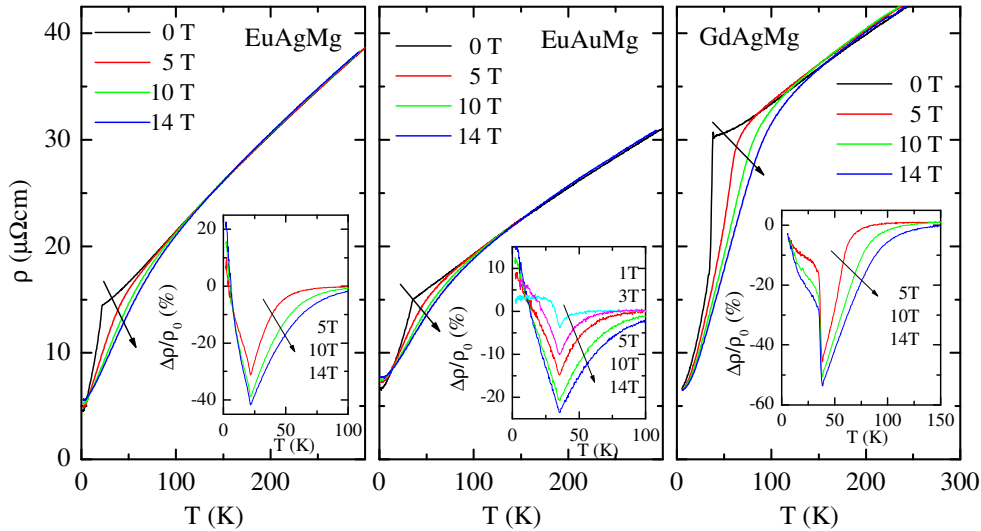}
\end{center}
\caption{\label{rhofm} Resistivity of ferromagnetic $RET$Mg for
different magnetic fields (increasing field strength is indicated
by the arrows). The insets show the magnetoresistance $\Delta
\rho/\rho_0= [\rho(H)-\rho(H=0)]/\rho(H=0)$.}
\end{figure}

In figure\,\ref{rhofm} the resistivity of Eu$T$Mg with $T={\rm
Ag}$ and Au and of GdAgMg is shown, which undergo a transition to
a ferromagnetic order at $T_c\simeq 22$\,K, $\simeq 36$\,K, and
$\simeq 39$\,K, respectively. With increasing magnetic field we
find a suppression of $\rho$, which is most pronounced in the
temperature range around $T_c$. Again we show $\Delta
\rho/\rho_0$ in the insets of figure\,\ref{rhofm}. In a field of
14\,T the maximum values of $\Delta \rho/\rho_0$ range from about
-55\,\% for GdAgMg to about -25\,\% for EuAuMg at the respective
$T_c$. Whereas for higher temperature $\Delta \rho/\rho_0$
continuously approaches zero, we observe a sign change for
Eu$T$Mg below $T_c$. Around 5\,K $\Delta \rho/\rho_0$ reaches up
to +20\,\% and +15\,\% for $T={\rm Ag}$ and Au, respectively. For
GdAgMg such a sign change is not observed, but from the steep
slope of $\Delta \rho/\rho_0$ at 2\,K we suspect that it will
also occur at somewhat lower temperature. Qualitatively, the
behavior of $\Delta \rho/\rho_0$ can be explained as follows: At
the lowest temperature the magnetic moments are almost completely
ordered and the (magneto-)resistance of the ferromagnetic $RET$Mg
is comparable to that of the paramagnetic Yb$T$Mg with respect to
both, the absolute value $\rho(0\,{\rm T})$ and the increase in a
magnetic field. With increasing temperature ferromagnetic spin
waves are excited and the spontaneous magnetization decreases.
Therefore, the charge carriers may be scattered by magnetic
excitations and as a consequence $\rho(T)$ (in zero field)
increases much faster for the ferromagnets than for the
paramagnets. The strongest increase of $\rho(T)$ occurs close to
$T_c$. Although this scattering is present in the ferromagnets
above $T_c$, the difference between $\rho(T)$ of the ferromagnets
and the paramagnets decreases with further increasing
temperature, since for higher temperature scattering by phonons
becomes more and more dominant and one may expect that the
scattering by phonons is not too different for the various
compounds. In the temperature range around $T_c$ the
magnetization strongly increases with increasing magnetic field.
Thus, the magnetic scattering of charge carriers can be
effectively suppressed in that temperature range because the
localized moments become aligned by increasing the magnetic
field. With further increasing temperature thermal fluctuations
increase. Thus the large negative magnetoresistance decreases
again, because the influence of the magnetic field on the
magnetization decreases. Note that LaAgMg does not fit into this
explanation, since its $\rho $ is much larger than  $\rho$ of the
isostructural-structural Gd$T$Mg. It remains to be clarified why
$\rho$ of LaAgMg is so large although there is no magnetic
scattering present.

\begin{figure}[t]
\begin{center}
\includegraphics[width=0.87\textwidth]{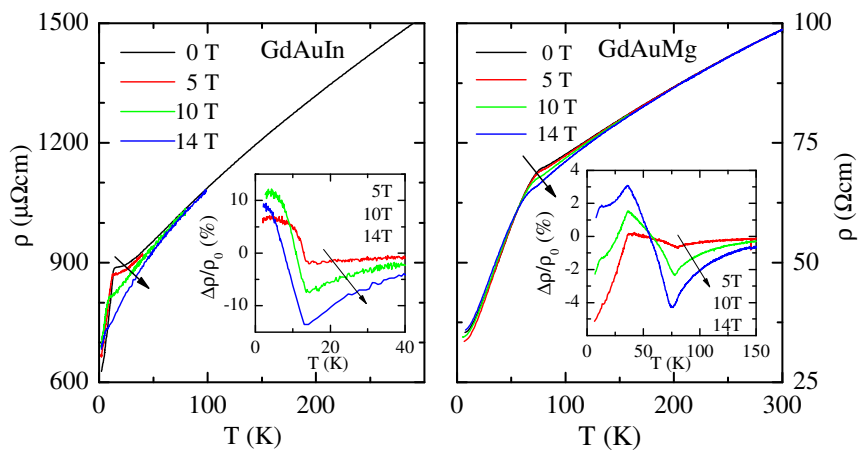}
\end{center}
\caption{\label{rhoafm} Resistivity of antiferromagnetic GdAuIn
and GdAuMg for different magnetic fields (increasing field
strength is indicated by the arrows). The insets show the
magnetoresistance $\Delta \rho/\rho_0=
[\rho(H)-\rho(H=0)]/\rho(H=0)$.}
\end{figure}

Figure\,\ref{rhoafm} displays the resistivity of GdAuIn and
GdAuMg, which order antiferromagnetically at $T_N\simeq 13$\,K and
$\simeq 81$\,K, respectively. Again we observe a negative
magnetoresistance in the temperature range around the ordering
temperature which can be traced back to the suppression of
magnetic fluctuations in a magnetic field. Compared to the
ferromagnetic compound, this effect is, however, much less
pronounced because the magnetic field tends to align the magnetic
moments parallel whereas the exchange favors an antiparallel
alignment. Despite the large absolute values of $\rho $ of GdAuIn
its magnetoresistance is rather strong. From the low $T_N$ of this
compound we conclude that the antiferromagnetic coupling is rather
weak and the field has a strong influence. For example, $T_N$
strongly shifts towards lower temperature with increasing field
and it seems that the enhanced resistivity due to magnetic
fluctuations around $T_N$ can be almost completely suppressed in
14\,T. We also observe a sign change of $\Delta \rho/\rho_0$ with
decreasing temperature, but at the lowest temperature $\rho$ does
not monotonously increase with field, in contrast to the other
$RET$Mg compounds (see figures\,\ref{rhopm} and\,\ref{rhofm}). We
think that this more complex behavior arises from the interplay
of the antiferromagnetic coupling and the strong magnetic field
which both suppress magnetic fluctuations but also act against
each other. A more detailed study down to lower temperature and
to higher magnetic fields is currently underway and will be
presented elsewhere. Since GdAuMg has the highest ordering
temperature of all the compounds studied, it will also have the
strongest (antiferromagnetic) coupling. Consequently, its
magnetoresistance is rather weak even in a field of 14\,T. Again,
there is a sign change of $\Delta \rho/\rho_0$ below $T_N$.
However, around $36$\,K $\Delta \rho/\rho_0$ shows a sharp kink
and a sign change of its slope. This kink of $\Delta \rho/\rho_0$
is present for all fields since it arises from a tiny kink of the
zero-field $\rho(T)$ curve at $\simeq 36$\,K, whereas the
$\rho(T)$ curves for larger fields do not show such an anomaly.
Additional anomalies also occur in our DC magnetization data
below about 50\,K and have been observed in AC susceptibility
measurements.\cite{P12} The additional anomalies in AC
susceptibility are most pronounced at 19\,K and have been
interpreted as evidence for a transition to a canted
antiferromagnetic phase at 19\,K. Thus, one may speculate whether
the kink in the zero-field $\rho(T)$ curve is related to a
precursor of such a spin-reorientation.

\subsection{Specific heat}

\begin{figure}[t]
\begin{center}
\includegraphics[width=0.87\textwidth]{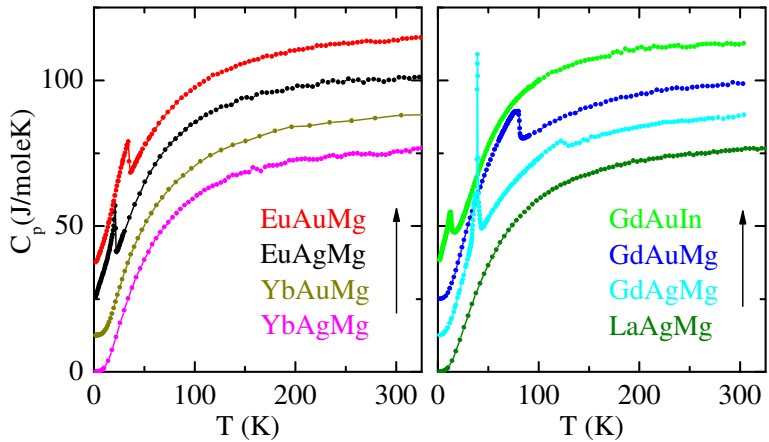}
\end{center}
\caption{\label{Cp} Specific heat of $RET$Mg and GdAuIn as a
function of temperature. In both panels the curves are shifted by
12.5\,J/moleK with respect to each other for the different
compounds as it is indicated by the arrows.}
\end{figure}

In figure\,\ref{Cp} we summarize the results of our specific heat
measurements for all compounds. As one may expect, the paramagnets
(YbAgMg, YbAuMg, and LaAgMg) show rather similar $C_p(T)$ curves
without any anomalies, since there are no phase transitions
(neither structural nor magnetic ones).\cite{remark} The
ferromagnets Eu$T$Mg ($T={\rm Ag}$, Au) and the antiferromagnets
GdMg$X$ ($X={\rm Au}$, In) show well-defined anomalies at the
respective Curie and N\'{e}el temperatures, respectively. In all four
cases the anomaly shape is typical for a (mean-field) second-order
phase transition without strong fluctuations. Above the respective
ordering temperature the $C_p(T)$ curves are again very similar
to each other and to those of the paramagnets. A clearly
different behaviour is observed for GdAgMg. Here, the anomaly of
$C_p$ at $T_c\simeq 39.5$\,K is very narrow and much higher than
in the other compounds. This anomaly resembles more the
$\lambda$-shape of a second-order phase transition with strong
fluctuations or to a broadened $\delta$-anomaly of a weakly
first-order phase transition. In addition, there is another clear
anomaly of $C_p(T)$ around 125\,K, whose shape is again typical
for a mean-field second-order phase transition.

In order to further investigate the origin of these anomalies we
analyze the change of (magnetic) entropy at the respective phase
transitions. For both, the Eu and the Gd compounds we expect that
the magnetic entropy will be dominated by the half-filled $4f$
shells with $S=7/2$ of Eu$^{2+}$ or Gd$^{3+}$. In addition, there
may be an additional contribution from a partial polarization of
the $5d$, $6s$ and $6p$ valence bands, but this will be much
smaller than the $4f$ contribution of
$S_{mag}=N_Ak_B\,\ln(2S+1)\simeq 17.3$\,J/moleK ($N_A$ and $k_B$
denote Avogadro's and Boltzmann's constant, respectively). We
assume that the total entropy consists of the sum of magnetic,
phononic and electronic contributions, i.\,e.\

\begin{equation}
S=S_{mag}+S_{ph}+S_{el}\,\,.
\end{equation}

\begin{figure}[t]
\begin{center}
\includegraphics[width=0.87\textwidth]{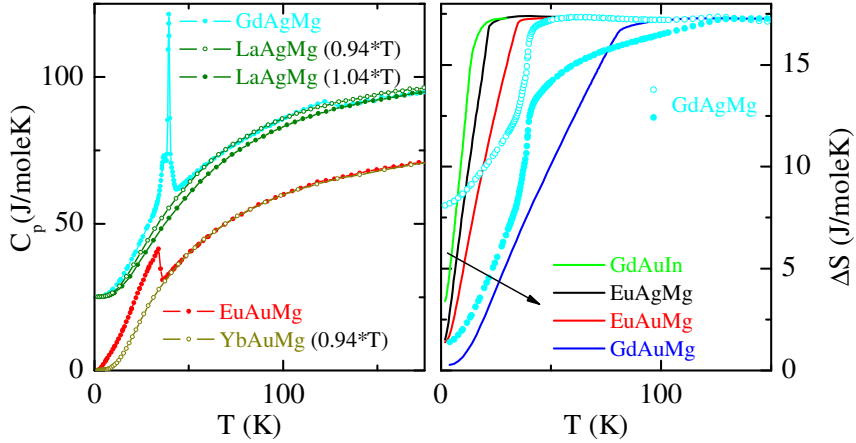}
\end{center}
\caption{\label{Smag} Left: Specific heat of $RE$AgMg for
magnetically ordering $RE={\rm Eu}$, Gd and for paramagnetic
$RE={\rm Yb}$, La. For the paramagnets the temperature has been
rescaled (the factors $a$ are given in the brackets) in such a
way that the $C_p(a*T)$ curves allow to estimate the phononic
contributions $C_{ph}(T)$ for $RE={\rm Eu}$ and Gd. Right:
Entropy changes due to the phase transitions obtained via an
integration of the differences $[C_p(T)-C_{ph}(T)]/T$ for the
different compounds. As expected for the half-filled $4f$ shell
the magnetic entropy $S_{mag}=N_Ak_B\,\ln(2S+1)$ with $S=7/2$ is
approached in most cases (lines). For GdAgMg (symbols) this is,
however, only the case when the entropy changes due to both
transitions are considered (see text).}
\end{figure}

In order to separate the magnetic  ($C_{mag}$) from the sum of
the phononic ($C_{ph}$) and the electronic contribution
($C_{el}$) we estimate $C_{ph}+C_{el}$ of the
(anti-)ferromagnetic compounds by using the measured $C_p(T)$ of
the isostructural paramagnets, for which $C_{mag}=0$. As a
typical example, we show the analysis for EuAuMg in the right
panel of figure\,\ref{Smag}, where $C_p(T)$ of YbAuMg has been
used to estimate $C_{ph}(T)$ of EuAuMg. For this estimate we
rescale the temperature axis of $C_p^{YbAuMg}$ until it agrees to
the measured $C_p^{EuAuMg}$ in the temperature range above $T_c$.
In this particular case a 6\% rescaling $T'=0.94\cdot T$ causes a
very good agreement up to the highest temperature (even at 300\,K
the curves deviate by less than 1\%). The entropy change due the
phase transition is then calculated from the difference $\Delta
C_p(T) = C_p^{EuAuMg}(T) - C_p^{YbAuMg}(T')$ via integration
$\Delta S(T) = \int \Delta C_p(T')/T' dT'$. In all cases the
integration constant has been chosen such that above the phase
transition $S_{mag}\simeq 17.3$\,J/moleK is reached as it is
expected for an $S=7/2$ system.

The right panel of figure\,\ref{Smag} shows the calculated entropy
changes. We find that $S_{mag}(T)$ strongly decreases below the
ordering temperature and roughly approaches zero for vanishing
temperature for all compounds except for GdAgMg. This is clear
evidence that the magnetic phase transitions can be viewed as an
ordering of a system with localized $S=7/2$ moments. In other
words, the hybridization of the $4f$ states with the valence
bands and/or a polarization of the valence bands hardly influence
$S_{mag}$. For GdAgMg the situation is more complex. Here, the
entropy change connected with the proposed magnetic ordering at
$T_c \simeq 39.5$\,K amounts only to about 50\,\% of the expected
$17.3$\,J/moleK. In the right panel of figure\,\ref{Smag} this is
shown by the open symbols, which are obtained when we estimate
the phononic background $C_{ph}^{GdAgMg}$ by scaling
$C_p^{LaAgMg}$ such that it coincides with the measured
$C_p^{GdAgMg}$ above about 40\,K (see upper curves in the left
panel of figure\,\ref{Smag}). However, it is also possible to
scale $C_p^{LaAgMg}$ such that it fits the measured
$C_p^{GdAgMg}$ only above the upper transition, i.\,e.\ above
about 130\,K. In this case the difference between both curves
yields the sum of the entropy changes due to both transition. As
it is shown by the full symbols in the right panel of
figure\,\ref{Smag} the total entropy change is again close to the
expected $S_{mag}\simeq 17.3$\,J/moleK. This indicates that for
GdAgMg a partial ordering of the magnetic moments sets in already
below $T_c^1\simeq 125$\,K and the complete ordering is obtained
only below $T_c^2\simeq 125$\,K. We have also performed DC
magnetization measurements as a function of temperature on that
sample, but we could not resolve any anomaly around 125\,K.
However, some small additional features, which start below about
120\,K, have been observed in AC susceptibility
measurements.\cite{P10}

\begin{figure}[t]
\begin{center}
\includegraphics[width=0.87\textwidth]{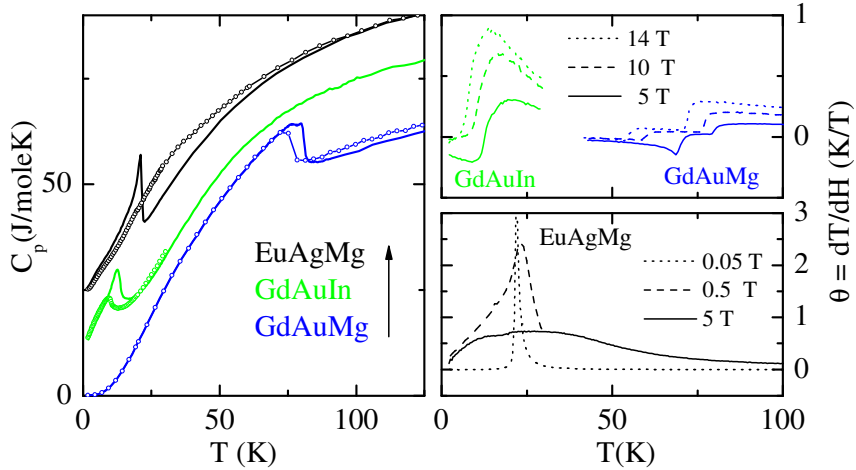}
\end{center}
\caption{\label{MKE} Left: Specific heat in zero (--) and a finite
magnetic field ($\circ$) of 5\,T (EuAgMg) or 10\,T (GdAuIn and
GdAuMg). For clarity the curves of the different compounds are
shifted with respect to each other. Right: Differential
magnetocaloric effect $\Theta = \frac{\partial T}{\partial H}$ of
the different compounds for various magnetic-field strengths (see
text).}
\end{figure}

The specific heat of the paramagnetic La- and Yb-based compounds
does not show any measurable magnetic-field dependence. For the
antiferromagnetic materials GdAuIn and GdAuMg the specific-heat
anomaly is systematically shifted towards lower temperature with
increasing field. As shown in the right panel of
figure\,\ref{MKE}, $T_N$ decreases from $\simeq 13.5$\,K to
$\simeq 10$\,K for GdAuIn in a field of 14\,T, and from $\simeq
81$\,K to $\simeq 78$\,K for GdAuMg. In figure\,\ref{MKE} we also present the
magnetic-field influence on $C_p$ for the ferromagnet EuAgMg.
Here, the transition is almost completely smeared out already in
a field of 5\,T. This is typical for ferromagnets because larger
fields cause a strong magnetization already well above the
zero-field $T_c$ and the magnetization does not develop
spontaneously below a critical temperature anymore. Strictly
speaking, a ferromagnetic transition temperature can only be defined
for zero magnetic field.

The strong magnetic-field influence on $C_p$ around the transition
temperature means that the (magnetic) entropy is strongly field
dependent. Thus, these materials are of potential interest with
respect for cooling or heating by adiabatic (de-)magnetization.
Usually only the magnetic entropy is magnetic-field dependent and
for an effective cooling $S_{mag}$ should therefore be large,
which is the case in our samples due to the large spin of 7/2 of
the half-filled $4f$ shell. The magnetocaloric effect, i.\,e.\
the magnetic-field induced variation of the sample temperature
under adiabatic conditions, can be either directly measured or it
can be calculated from measurements of specific heat and
magnetization via the thermodynamic relation
\begin{equation}
\Theta = \frac{\partial T}{\partial H} =  - \frac{T}{C}
\left.\frac{\partial M}{\partial T}\right|_H \, . \label{mke}
\end{equation}
Equation\,\ref{mke} yields an expression for the differential
magnetocaloric effect $\Theta $ as a function of temperature for
a fixed magnetic field and the field-induced temperature change
$\Delta T =T(H_1)-T(H_2)$ can be obtained by numerical
calculation from measurements in constant magnetic fields with
$H_1\le H \le H_2$.

The upper right panel of figure\,\ref{MKE} shows the differential
magnetocaloric effect of the antiferromagnetic compounds. In a
field of 5\,T we find for GdAuIn a small positive $\Theta \simeq
0.3$\,K/T above and $\simeq -0.2$\,K/T below $T_N$, and $\Theta $
vanishes towards both, higher and lower temperatures. This
behavior is expected via equation~\ref{mke} from the typical
behavior of an antiferromagnet. With decreasing temperature
$-\partial M /\partial T$ increases according to a Curie-Weiss
law and $C_p$ decreases. Both effects cause an increase of
$\Theta $ until $T_N$ is reached. For low fields the
magnetization usually changes its slope at $T_N$, and $\Theta$
its sign. For $T\rightarrow 0$\,K the magnetization usually
approaches a constant value, i.\,e.\ $\Theta $ is expected to
vanish. For fields lower than 5\,T the temperature dependencies
of the $\Theta (T)$ curves are very similar to that of $\Theta
(T)$ in $H=5$\,T. However, the absolute values of $\Theta (T)$
are reduced, because in low fields the magnetization of an
antiferromagnet is roughly proportional to the applied field,
whereas $C_p/T$ does not change too much. Therefore, the
proportionality $\Theta \propto H$ is roughly fulfilled for
$H<5$\,T. For higher fields the maximum of $\Theta $ further
increases, but $\Theta $ remains positive, because the sign
change of $\partial M /\partial T$ is suppressed in large enough
fields. For example, a field of 14\,T causes a low-temperature
magnetization of more than 80\,\% of the saturation moment of
7\,$\mu_B$/f.u. of the half-filled $4f$ shell. For GdAuMg one
would expect a qualitatively similar behavior as observed in
GdAuIn, but with smaller absolute values of $\Theta$ due to the
larger $T_N$. This is, however, not the case. The main difference
is that $\Theta$ only shows a step-like decrease close to $T_N$,
whereas the sign change occurs about 10 to 15\,K below $T_N$.
This more complex behavior points towards a magnetic-field
dependent reorientation of the magnetic moments. As already
mentioned above, evidence for such a behavior has also been found
from measurements of low-field AC susceptibility and M\"{o}ssbauer
spectroscopy.\cite{P12}

The lower right panel displays the differential magnetocaloric
effect for ferromagnetic EuAgMg. In the lowest field of 50\,mT we
find a very sharp peak with $\Theta \simeq 3$\,K/T, which is
located close to the zero-field $T_c$. With increasing field the
peak rapidly decreases in height and strongly broadens. For
example, $\Theta$ lies between 0.6 and 0.7\,K/T over a
temperature range from about 10 to 40\,K in a field of 5\,T. Such
a behavior is typical for a ferromagnet because the transition
strongly smears out with increasing field. Due to the large spin
the magnetocaloric effect of EuAgMg is relatively large. However,
larger effects are found e.\,g.\ in some $RM_2$ compounds with
$R={\rm Dy}$, Ho, Er and $M={\rm Co}$, Ni, Al. Here, larger
magnetic entropy changes occur due to the larger $J=L+S$ values
of $R$, additional crystalline electric field effects, and
contributions from magnetic
Co.\cite{rankePRB98,rankePRB01,oliveiraPRB02} Moreover, a
so-called giant magnetocaloric effect observed in
Gd$_5$(Si$_2$Ge$_2$) has attracted a lot of
attention.\cite{pecharskyPRL97} This giant magnetocaloric effect
is related to a second-order transition from a para- to a
ferromagnetic(I) state, which then transforms via a first-order
transition to a ferromagnetic(I) state.

\section{Summary}

We have performed a systematic study of the resistance and the
specific heat as a function of temperature (${\rm 2\,K} <T<{\rm
300\,K}$) and magnetic field (up to 14\,T) on a series of
rare-earth ($RM$) transition-metal $(T)$ Mg compounds $RET$Mg and
on GdAuIn. For $RE = {\rm La}$ (Yb) the $4f$ shell is empty
(filled) and the respective compounds are paramagnetic metals,
whereas the Eu- and Gd-based materials show ferro- or
antiferromagnetic ordering arising from the half-filled $4f$
shell of Eu$^{2+}$ or Gd$^{3+}$. All compounds show a metallic
resistivity, but the absolute values of $\rho$ vary over three
orders of magnitude from a few $\mu\Omega$cm in YbAuMg to
$m\Omega$cm in GdAuIn. The absolute value  of $\rho$ increases
with increasing number of valence electrons, what may arise from
the different crystal structures of the di- (Yb,Eu) and trivalent
(La,Gd) $RE$-based compounds. Both, the ferromagnetic and the
antiferromagnetic ordering causes an abrupt decrease of $\rho$
below the respective Curie- and N{\'e}el-temperature, because the
scattering of charge carriers on magnetic excitations freezes out
in the ordered phase. For the ferromagnets we find a large
negative magnetoresistance, which is most pronounced  close to
$T_c$, where a decrease of $\rho$ of order -5\%/T is observed for
fields below 5\,T. As expected, a significantly smaller decrease
of $\rho$ (of order -0.2\%/T) is found for the antiferromagnets
close to $T_N$. With decreasing temperature the magnetoresistance
becomes weaker and for most of the magnetically ordered compounds
it is positive at the lowest temperature and comparable to the
positive magnetoresistance of the paramagnets. For all ferro- and
antiferromagnetic compounds the magnetic ordering causes
well-defined anomalies in the specific heat. Our analysis of the
related entropy change clearly shows that the magnetic
transitions are well described by an ordering of localized
$S=7/2$ moments of the half-filled $4f$ shell of Eu$^{2+}$ or
Gd$^{3+}$. Additional effects, e.\,g. from a polarization of the
valence bands or a hybridization between 4f states and the
valence bands, play a minor role. However, for GdAgMg we find
clear evidence for two transitions at 125 and 39\,K. In
susceptibility measurements a sizeable spontaneous magnetization
is only found below 39\,K, but the entropy change due to this
transition is only about half as large as expected for an $S=7/2$
system. The expected magnetic entropy change is achieved when the
sum of the entropy changes due to both transitions is considered.
Thus, the specific heat data suggest that the complete magnetic
order is reached via two transitions. This more complex behavior
of GdAgMg is not yet understood and deserves further
investigations. For both, the ferro- and the antiferromagnetic
compounds we observe a sizeable magnetocaloric effect $\Theta$,
which is most pronounced for the ferromagnets in low fields and
close to $T_c$. For EuAgMg a relatively large $\Theta \ge
0.6$\,K/T is present over a large field and temperature range, but
this value still remains well below those observed in various
other rare-earth based materials.

\ack

We acknowledge valuable discussion with L.\,H.\,Tjeng. We thank
the Degussa-H\"{u}ls AG for a generous gift of noble metals. This work
was supported by the Deutsche Forschungsgemeinschaft through the
priority programme SPP 1166 "Lanthanoidspezifische
Funktionalit\"{a}ten in Molek\"{u}l und Material".


\begin{thebibliography}{99}
\bibitem{remarkLa} For the La-based compounds we cannot show the
Kolher plot, because we have performed only one field dependent
$\rho$ measurement ($\rho $\,(14\,T) of LaAgMg).
\bibitem{remark} For YbAuMg there is a very tiny anomaly in $C_p(T)$
around 2.3\,K, which cannot be resolved in figure\,\ref{Cp}. This
anomaly is suppressed in a magnetic field of 3\,T and most
probably arises from a small amount of an impurity phase of
Yb$_2$O$_3$.
\end{thebibliography}
\end{document}